\documentclass{article}
\usepackage{graphics,pstricks,pst-node}
\newcommand{\EcoLab}{{\sffamily\slshape
    \mbox{\raisebox{.5ex}{Eco}\hspace{-.4em}{\makebox[.5em]{L}ab}}}}

\title{Rationality in the Theory of the Firm}
\author{Russell K. Standish, \\
Mathematics and Statistics, University of New South Wales \and Stephen L.
Keen \\
Economics and Finance, University of Western Sydney }

\begin{document}
\maketitle

\begin{abstract}
 We have previously presented a critique of the standard Marshallian
theory of the firm, and developed an alternative formulation that
better agreed with the results of simulation. An incorrect
mathematical fact was used in our previous presentation. This paper
deals with correcting the derivation of the Keen equilibrium, and
generalising the result to the asymmetric case. As well, we discuss
the notion of rationality employed, and how this plays out in a two
player version of the game.


\end{abstract}



\section{Introduction}

Keen \cite[Ch 4]{Keen02} pointed out a fundamental flaw with the standard
Marshallian theory of the firm, whereby the market demand function
$P(Q)$ (price of a good given total market production Q) is assumed to
be a decreasing function of $Q$ (i.e. $P'(Q)<0$), yet at the same time, for
a large number of firms, each individual firm's production $q_i$ has
no effect on market price, ie $\partial P/\partial q_i=0$. Yet it is
easy to see from elementary calculus, that these two conditions cannot be true
simultaneously, as first noted by Stigler~\cite{Stigler57}.

Marshallian analysis proceeds under this assumption that individual
firms' actions have no effect on the market, leading to the profit
maximum for each firm to occur when it's marginal cost is equal
to the market price.

In \cite{Standish-Keen04, Keen-Standish06}, we argue that the
economy's equilibrium will not occur at the zero of the partial derivative of
the individual profit function, but rather when the {\em total}
derivative of each individual profit with respect to market production
is simultaneously satisfied. This leads to a revised prediction of the
difference between market price and marginal cost being related to the
slope of the demand curve:
\begin{equation}\label{KeenEq}
P(Q)-MC(q_i) = -nq_i P'(Q)
\end{equation}

Furthermore, a simple reactive rational agent model of the firm
produced results compatible with the Keen equilibrium, and not the
Cournot-Nash equilibrium predicted from standard Marshallian
analysis. It should be pointed out that this agent model makes neither
the partial derivative assumption of Marshallian analysis, nor the
total derivative assumption of Keen's analysis, but rather the agents
seek to always optimise their profits assuming the past is a guide to
the future.

Anglin~\cite{Anglin08} critiqued our 2006 paper, but the critique was
not without its own mathematical difficulties. We extensively rebutted
his paper in a submission to the same journal in which our 2006 paper,
and Anglin's critique appeared. This was rejected on editorial
privilege. We chose not to publish the rebuttal in another journal, as
the rebuttal doesn't advance the state of the field, but have made it
available via arXiv \cite{Standish-Keen-Anglin-rebuttal}, for thos who
might be interested.

Nevertheless, in the course of corresponding with Anglin, the main
issue bothering Anglin was identified as an erroneous mathematical
assumption we made for the value of $dq_i/dQ$, for which no such
assumption can be made. This paper serves to correct
the analysis, and also correctly generalise the Keen analysis to the
asymmetric firm case. As a consequence, our previous attempt described
in section 3 of \cite{Keen-Standish06}, which Anglin ridiculed as
``conjectural variation'', is no longer relevant. 

The purpose of this paper is not to rebut Anglin's paper, but to
correct a problematic assertion in our work, and consequently extend
the Keen analysis to asymmetric firm response.

\section{The profit formula}

We take as our starting point, the usual profit formula of a single
product market with $n$ firms:
\begin{equation}\label{profit}
\pi_i = q_iP(Q)-\int_0^{q_i} MC(q_i) dq_i,
\end{equation}
where $\pi_i$ is the profit obtained by firm $i$, as a function of its
production $q_i$, and the total market production $Q=\sum_i q_i$. The
function $P(Q)$ is the demand curve, namely the price the good
achieves when $Q$ items of the good is available on the market. In the
following, $P(Q)$ is taken to be a monotonically decreasing curve
($P'<0$). The function $MC(q_i)$ is the marginal cost of producing an
extra item of the good, given that a firm is producing $q_i$ items.

\section{Rationality}

The key concept of the {\em rational agent}, or {\em homo economicus}
is that the agent chooses from an array of actions so as to maximise
some utility function. In the context of the theory of the firm, the
utility functions are given by $\pi_i$ in eq (\ref{profit}), and the
choices are the production values chosen by the individual firms.

Intrinsic to the notion of rationality is the property of
determinism. Given a single best course of action that maximises
utility, the agent must choose that action. Only where two equally
good courses of action occur, might the agents behave
stochastically. This deterministic behaviour of the agents is the key
to understanding the stability of the Keen equilibrium, and the
instability of the Cournot equilibrium, which is the outcome of traditional
Marshallian analysis.

When setting up a game, it is important to circumscribe what information
the agents have access to. Clearly, if the agents know what the total market
production $Q$ will be in the next cycle, as well as their marginal
cost $MC$, the rational value of $q_i$ can be found by setting the
partial derivative of $\pi_i$ to zero: 
\begin{equation}\label{dpidq}
\frac{\partial\pi_i}{\partial q_i} = P+q_i\frac{\partial P}{\partial q_i}-MC(q_i) = 0.
\end{equation}

Indeed, the Marshallian theory further assumes that in the limit as
the number of firms $n$ tends to infinity, the term $\partial
P/\partial q_i\rightarrow0$, to arrive at the ultimate result that
price will tend to the marginal cost (assuming a unique marginal cost
exists over all firms)\cite[p. 322]{MasColell}. This assumption is
strictly false, as shown by \cite{Stigler57}. Instead, $\partial
P/\partial q_i = dP/dQ$, which is independent of the number firms in
the economy. The Cournot-Nash model starts with each agent knowing
that all other agents are rational, and their marginal cost curves,
consequently (under the right circumstances) being able to predict the
optimal production levels for each agent. Therefore, the total
production $Q$ is predictable, and equation (\ref{dpidq}) should
hold. Furthermore, for certain distributions of market share (eg the
symmetric case of equal market share where $q_i=Q/n$), individual production
levels vanish in the limit $n\rightarrow\infty$. Therefore
$P\rightarrow MC$. This result is known as the Cournot theorem.

However, it is completely unrealistic for the firms to be able to
predict market production (and hence price). Firms cannot know whether
their competitors will act completely rationally, and details such as
the marginal cost curve for each firm, and even the total number of
players is unlikely to be known. So equation (\ref{dpidq}) cannot be 
correct. Instead, firms can really only assume that the price tomorrow
will most likely be similar today, and that the best they can do is incrementally adjust
their output to ``grope for'' the optimal production value. So in our
model, firms have a choice between increasing production or decreasing
it. If the previous round's production change caused a rise in profits,
the rational thing to do is to repeat the action. If, on the other
hand, it leads to a decrease in profit, the opposite action should be
taken. At equilibrium, one would expect the production to be
continuously increased and decreased in a cycle with no net movement.

\section{Game theoretic analysis of the Cournot equilibrium}\label{GameTheoretic}

For simplicity, assume a two firm system with identical constant
marginal costs, that has been initialised at its Cournot equilibrium
($P+QP'/2-MC=0, q_1=q_2=Q/2$). There are three possible outcomes for
the next step:
\begin{enumerate}
\item \label{both-increase} Both firms increase production. This reduces the price fetched
for each firm. The right hand side of equation \ref{dpidq} becomes
negative, reducing the profits of both firms. The logical next step is
for both firms to decrease production, which is covered under item \ref{both-decrease}

\item One firm increases production whilst the other decreases it. If
the production increment is the same in each case, then the market
price does not change. The net effect is of one
firm gaining market share at the expense of the other. In this case,
the firm losing market share will switch to increasing production,
whilst the other firm will continue increasing production. This is
situation described by item \ref{both-increase}.

If the production increments differed between firms, then there are
two cases: if the firm with larger increment increases, and the
increment is sufficiently big, then profit levels will fall for both
firms. The dynamics returns to the original (Cournot)
point. Otherwise, the firm losing market share will switch to
increasing, which is situation \ref{both-increase}.

\item \label{both-decrease} 
Both firms decrease production. In this case, provided the price
is higher than the monopoly price, both firms' profits will rise,
leading to another round of production decreases.

\end{enumerate}

The net result is that the Cournot equilibrium is unstable in the
direction of both firms decreasing production. The n-firm case can be
analysed in the same way \cite{Standish-Keen04}. The situation where
the majority of firms are decreasing their production simultaneously
will occur by chance within a few cycles of the system
initialisation. From there, the entrainment of all firms into the
production-reducing behaviour happens rapidly, until the system
stabilises at monopoly prices. 

It is important to note, that this effect depends on the deterministic
nature of the agent behaviour, a result of the assumption of
rationality. Presumably, most real economic agents are not as rational
as this, and the introduction of 30\% irrationality into the agents is
sufficient to ensure competitive pricing \cite{Standish-Keen04}.

\section{Derivation of the Keen equilibrium}\label{keen-eq-deriv}

Mathematically, global equilibrium will occur when all partial
derivatives $\partial\pi_i/\partial q_j$ vanish. However, this
situation can never pertain, as $\forall i\ne j,\, \partial\pi_i/\partial q_j = q_iP' <
0$, except possibly for the trivial solution $Q=0$.

Instead we propose the condition that all firm's profits are maximised
with respect to total industry output $d\pi_i/dQ=0$. This constrains
the dynamics of firms' outputs to an $n-1$-dimensional polyhedron, but
otherwise does not specify what the individual firms should do. As an
equilibrium condition, it is vulnerable to a single firm ``stealing''
market share. However, no firm acts in isolation. The other firms will
react, negating the benefit obtained by first firm, causing the system
to settle back to the  $d\pi_i/dQ=0$ manifold.

The derivation of the Keen equilibrium follows the presentation in
\cite{Keen-Standish06}. The total derivative of an individual firm's
profit is given by
\begin{equation}  \label{max-profit}
\frac{d\pi_i}{dQ} = P \frac{dq_i}{dQ} + q_i \frac{dP}{dQ} -
MC(q_i)\frac{dq_i}{dQ}
\end{equation}
which is zero at the Keen equilibrium.

In terms of the model introduced in \S\ref{GameTheoretic}, there is no
absolute equilibrium, but rather a limit cycle where the individual
firms are ``jiggling'' their outputs around the equilibrium value. If
we average over this limit cycle, and retaining only zeroth order
terms in $\Delta q_i$, we get
\begin{eqnarray}\label{Keen-eq}
\left\langle\frac{d\pi_i}{dQ}\right\rangle &=& 
P(\langle Q\rangle) \left\langle\frac{dq_i}{dQ}\right\rangle +
\langle q_i \rangle P'(\langle Q\rangle) - MC(\langle q_i \rangle)
\left\langle\frac{dq_i}{dQ}\right\rangle \nonumber\\
&=& P\theta_i + q_i P' - MC(q_i)\theta_i = 0. 
\end{eqnarray}
where $\theta_i=\langle dq_i/dQ\rangle$ and the terms $P$, $P'$ and
$q_i$ refer to the equilibrium average values of these quantities. The
$\theta_i$ terms are normalised: $\sum_i\theta_i = dQ/dQ = 1$. They
can be considered to be the (normalised) {\em responsiveness} of the
firms to changing market conditions

The symmetric firm case corresponds to setting $\theta_i=1/n$, which
leads to the equation: \begin{equation}\label{Orig-Keen-eq} P+nq_iP' -
MC(q_i) = 0 \end{equation} which is equation (6) of
\cite{Keen-Standish06}.

In our previous expositions \cite{Keen-Standish06,Standish-Keen04}, we
incorrectly set $dq_i/dQ=\sum_j\partial q_j/\partial q_i = 1$, which
as pointed out in a critique by Paul Anglin\cite{Anglin08}, when
coupled with $q_i(Q=0)=0$ leads to the unjustifiable conclusion that
$q_i = Q/n$ at all times. Now, in equation (\ref{Keen-eq}), the values
$\theta_i$ only refer to the derivatives at equilibrium, so there is
no necessity for (\ref{Orig-Keen-eq}) to entail an equi-partition of
the market share.

\section{Testing the Keen equilibrium}

We can use an agent-based computational model based on the model
introduced in \S\ref{GameTheoretic} to test the Keen equilibrium, or
more specifically, equation (\ref{Keen-eq}). The terms $P(Q)$, $q_i$
and $MC_i$ are all available as part of the model. In addition, each
agent has an attribute $\delta_i$, which is the amount that agent $i$
varies its production up or down each time step.

We can compute the quantity $\theta_i$ by averaging $\Delta q_i/\Delta
Q$ over $\theta_w = 10$ time steps.

In the following experiment with 1000 firms, we used a linear pricing
function $P=11-Q/3$, and constant marginal costs $MC_i$ drawn
uniformly from the range $[0.5,0.5)$. The $\delta_i$ increments were
drawn from a half normal distribution --- ie the absolute values of
normally distributed deviates with zero mean and standard deviation
$0.002/n$. The code implementing the model, and the experimental
parameter script is available as firmmodel.1.D7, from the \EcoLab{}
website (http://ecolab.sf.net).

\begin{figure}
\begin{center}
\resizebox{\textwidth}{!}{\includegraphics{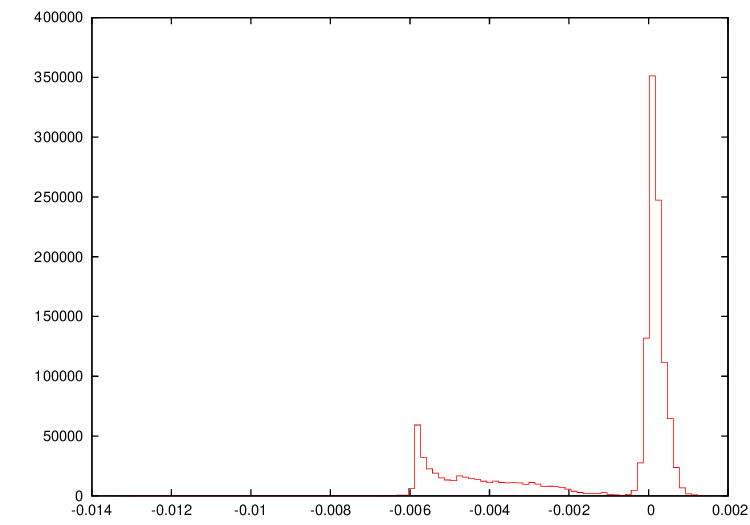}}
\end{center}
\caption{Histogram of the statistic $q_iP'+\theta_i(P-MC_i)$ for 1378
replications of the 1000 firm model. Most of the values are clustered
close to zero, but a sizable minority have non-zero negative
values. The left hand half of the histogram also has non-zero bins,
but are not visible with the shown vertical scale }
\label{keen-cond-histo}
\end{figure}

Figure \ref{keen-cond-histo} shows the histogram of values taken by
the statistic $q_iP'+\theta_i(P-M_i)$ for the 1000 firms over 1378
replications. The vast majority of observed values are consistent with
zero, the predicted value of eq (\ref{Keen-eq}), however there is a
significant minority of outliers, which are not explained within the
theory presented in \S\ref{keen-eq-deriv}. 

\begin{figure}
\begin{center}
\resizebox{\textwidth}{!}{\includegraphics{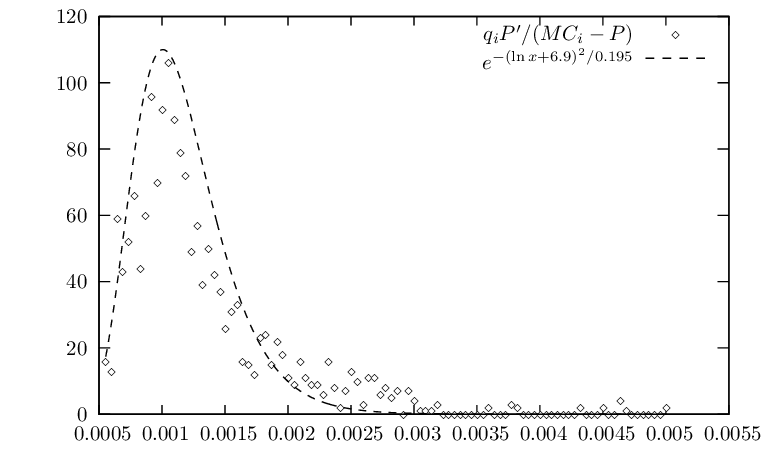}}
\end{center}
\caption{Histogram of the statistic $q_iP'/(MC_i-P)$ for a single run
of the 1000 firm model. The dashed line shows a fitted log-normal distribution.}
\label{cournot-histo}
\end{figure}

We can, however, consider the statistic $\xi=q_iP'/(M_i-P)$, which in the
Cournot theory should be one by (\ref{dpidq}). Figure
\ref{cournot-histo} shows a histogram of $\xi$ for a single run of the
1000 firm model. The values approximately fit a lognormal
distribution, from which we can see that value $\xi=1$ lies some 35
``sigmas" to the right of the mean, ie the Cournot prediction
(\ref{dpidq}) is excluded to the tune of $p\approx 10^{-267}$.

\section{Conclusion}

In this paper, we have discussed the behaviour of an n-player game of
rational, but not clairvoyant, agents. This exhibits a phase of
coordinated behaviour of the agents that brings market prices to
near monopoly levels due to the very rationality of the
agents rather than any overt coordination mechanism. We use numerical
simulations to reject the traditional Cournot-Nash solution of the game.

\bibliographystyle{plain}

\bibliography{rus}

\end{document}